\documentclass[nofootinbib,preprint,aps,draft,pre,superscriptaddress,citeautoscript,floatfix]{revtex4-2}
\setcitestyle{super}
\usepackage{graphics}
\usepackage{graphicx}
\usepackage{mathrsfs}
\usepackage{amsmath}
\usepackage{bm}
\usepackage{color}
\usepackage{float}

\begin{document}
%%%%%%%%%%%%%%%%%%%%%%%%%%%%%%%%%

\title{The Chain Flexibility Effects on the Self-assembly of Diblock Copolymer in Thin Film} 
\author{Mingyang Chen}
\affiliation{Center of Soft Matter Physics and its Applications, Beihang University, Beijing 100191, China}
\affiliation{School of Physics, Beihang University, Beijing 100191, China}
\author{Yuguo Chen}
\affiliation{Center of Soft Matter Physics and its Applications, Beihang University, Beijing 100191, China}
\affiliation{School of Chemistry, Beihang University, Beijing 100191, China}
\author{Yanyan Zhu}
\affiliation{Center of Soft Matter Physics and its Applications, Beihang University, Beijing 100191, China}
\affiliation{School of Physics, Beihang University, Beijing 100191, China}
\author{Ying Jiang}
\email{yjiang@buaa.edu.cn}
\email{andelman@tauex.tau.ac.il}
\email{manxk@buaa.edu.cn}
\affiliation{Center of Soft Matter Physics and its Applications, Beihang University, Beijing 100191, China}
\affiliation{School of Chemistry, Beihang University, Beijing 100191, China}
\author{David Andelman}
\email{yjiang@buaa.edu.cn}
\email{andelman@tauex.tau.ac.il}
\email{manxk@buaa.edu.cn}
\affiliation{School of Physics and Astronomy, Tel Aviv University, Ramat Aviv 69978, Tel Aviv, Israel}
\author{Xingkun Man}
\email{yjiang@buaa.edu.cn}
\email{andelman@tauex.tau.ac.il}
\email{manxk@buaa.edu.cn}
\affiliation{Center of Soft Matter Physics and its Applications, Beihang University, Beijing 100191, China}
\affiliation{School of Physics, Beihang University, Beijing 100191, China}
\affiliation{ Peng Huanwu Collaborative Center for Research and Education, Beihang University, Beijing 100191, China}

%%%%%%%%%%%%%%%%%%%%%%
\begin{abstract}
%\section*{abstract}
%%%%%%%%%%%%%%%%%%%%%%

We investigate the effects of chain flexibility on the self-assembly behavior of symmetric diblock copolymers (BCPs) when they are confined as a thin film between two surfaces. Employing worm-like chain (WLC) self-consistent field theory, we study the relative stability of parallel (L$_{\parallel}$) and perpendicular (L$_{\perp}$) orientations of BCP lamellar phases,  ranging in chain flexibility from flexible Gaussian chains to semi-flexible and rigid chains.  For flat and neutral bounding surfaces (no surface preference for one of the two BCP components), the stability of the L$_{\perp}$ lamellae increases with chain rigidity. When the top surface is flat and the bottom substrate is corrugated, increasing the surface roughness enhances the stability of the L$_{\perp}$ lamellae for flexible Gaussian chains. However, an opposite behavior is observed for rigid chains, where the L$_{\perp}$ stability decreases as the substrate roughness increases. We further show that as the substrate roughness increases, the critical value of the substrate preference, $u^{*}$,  corresponding to an L$_{\perp}$-to-L$_{\parallel}$ transition, decreases for rigid chains, while it increases for flexible Gaussian chains. Our results highlight the physical mechanism of tailoring the orientation of lamellar phases in thin-film setups. This is of importance, in particular, for short (semi-flexible or rigid) chains that are in high demand in emerging nanolithography and other industrial applications.

\end{abstract}

\maketitle

%%%%%%%%%%%%%%%%%%%%%%%%%%%%%%%%%%%%%%%%%%%%%%%%%%%%%%%
\section*{Introduction}
%%%%%%%%%%%%%%%%%%%%%%%%%%%%%%%%%%%%%%%%%%%%%%%%%%%%%%%

With the growth of the microelectronic industry, small-size transistors with well\,-controlled nanostructures extending over large length scales are in high demand~\cite{1Hillmyer2015}.
The self-assembly of block copolymers (BCPs) is considered a promising venue to meet these requirements~\cite{2Man2021,3Russell2000,4Park2003}. The simplest of all BCP architectures is the AB diblock copolymer (di-BCP), where each chain is composed of two chemically distinct blocks that are covalently tethered together. Di-BCPs can microphase separate to form a variety of equilibrium nanostructures such as lamellar and cylindrical phases~\cite{5Bates1994,6LWH2013}. By tuning the BCP polymerization index $N$ (or equivalently, the molecular weight)~\cite{7Bates1999,1Hillmyer2015} and the polymer chain rigidity $via$ the choice of the chemical constituents for monomers (e.g, conjugated polymers~\cite{LEE2015,Gu2019}), these spontaneously generated nanostructures have typical periodicities in the 5-100 nm range, which make them ideal for patterning technologies~\cite{8cheng2001,9Bates2014,Peng2020}.

A considerable amount of effort has been devoted to producing lamellae or cylinders of BCP thin-film of small domain size~\cite{10Jung2010,11chaudhari2014,12Wang2019,13Xu2021}. Using a poly(styrene-$b$-dimethylsiloxane) (PS-PDMS) block copolymer with a total molar mass of 16 kg/mol ($N$\,=\,$160$), Jung {\it {et al.}} reported~\cite{10Jung2010} formation of arrays of parallel cylinders with a periodicity of 17 nm. Later, a 6.4 kg/mol ($N$\,=\,$63$) poly(styrene-$b$-4vinylpyridine) (PS-P4VP) block copolymer was used to obtain lamellae with a characteristic size of 10.3 nm~\cite{11chaudhari2014}. Deng {\it {et al.}} reported~\cite{12Wang2019} a low molar mass poly(pentadecafluorooctyl methacrylate)-$b$-polyhydroxystyrene (PPDFMA-PHS) ($N$\,=\,$38$) forming nanodomains with a characteristic size of 9.8 nm. Recently, Xu {\it {et al.}} used~\cite{13Xu2021} poly(styrene-$b$-(4-vinylpyridine)propane-1-sulfonate) (PS-PVPS) with $N$\,=\,$21$ to obtain a lamellar phase with a periodicity of 5.7 nm.

For these BCPs with high Flory-Huggins interaction parameter $\chi$ associated with a low degree of polymerization~\cite{30Matsen1996,JiangPRE2011,38Jiang2013,JiangPRL2013,JeffChen2016_PPS},  the A/B constituent blocks usually have considerably different surface energies. However, for many materials and engineering applications, one has to rely on a thin-film set-up to produce BCP films with a perpendicular orientation of the BCP lamellae or cylinders with respect to the underlying substrate~\cite{9Bates2014}. For example, in optoelectronic applications, controlling the orientation of the lamellar phase confined in a thin film has attracted considerable attention because the chain alignment is closely related to the efficiency of charge transport~\cite{Gu2019, Peng2020}. Therefore, various techniques, such as solvent vapor annealing (SVA) and patterned substrates~\cite{14Aissou2015,15Suh2017,16Shin2011}, have been developed to eliminate the surface preference and to stabilize the perpendicular orientation (L$_{\perp}$).

Corrugated substrates are usually used to overcome such surface preference to obtain the perpendicular L$_{\perp}$ phase~\cite{2Man2021}. Theoretical studies based on the self-consistent field theory (SCFT) indicate that the parameter $q_sR$ plays the key factor in inducing an L$_{\parallel}$-to-L$_{\perp}$ phase transition~\cite{20Man2015,21Man2016}, where $q_s$ is the wavenumber and $R$ is the amplitude of a sinusoidally corrugated substrate. In yet another work~\cite{22Zhu2019}, similar results were found for the orientation of cylindrical BCP thin films placed on top of a corrugated surface. Employing dissipative particle dynamics (DPD) simulation~\cite{23Ross2020}, it was found that the BCP arrangement can induce an L$_{\perp}$ perpendicular orientation of the higher layer BCP in a uniform multilayer nano-system.  Furthermore, a phenomenological theory compared the L$_{\parallel}$ and L$_{\perp}$ free energies on corrugated substrates by using the analogy between smectic liquid crystals and lamellar BCP. The main finding was that by increasing the corrugation amplitude, the perpendicular lamellar (L$_{\perp}$) is the preferred phase~\cite{24Tsori2003,25Tsori2005,26Turner1992}. Such theoretical findings are consistent with experimental results~\cite{17Sivanniah2003,18Sivanniah2005, 19Kim2013, 14Aissou2015}.

Previous theoretical works have been focusing on the self-assembly of thin films of {\it flexible} chain BCPs. However, the above-mentioned short BCP chains and some other types of BCPs (like conjugated BCPs) whose semi-flexibility is particularly pronounced~\cite{27Porod1949,28Nicholas2020,29Scott2022, LEE2015,Gu2019,Peng2020}, no longer have the coiled conformation as was assumed in most previous theoretical works. Therefore, the flexible chain assumption based on the Gaussian chain (GSC) model is inappropriate to describe the chain statistics for polymers with a low degree of polymerization or chains that are not fully flexible~\cite{Matsen2012,31Jiang2016}. For such polymers, a more suitable model is the wormlike chain (WLC) model~\cite{27Porod1949,32Saito1967} since it facilitates the study of the conformational variations of polymer chains, which deviates from the GSC model. A WLC is commonly used to describe a semiflexible polymer where the polymer appears rigid approximately within the persistence length $\lambda$. Thus, the polymer chain conformations can be quantified by the ratio $L/\lambda$, where $L$ denotes the total contour length of a WLC. In the limit of $L/\lambda \gg 1$, the WLC model exactly recovers the GSC, where the effective Kuhn length is identified as $a=2\lambda$ and $L/a$ is equivalent to the degree of polymerization. On the other hand, for $L/\lambda \sim 1$ the WLC model crosses over and describes a rigid rod chain. Any finite $L/a$ ratio in the WLC model gives rise to a theory that contributes to the effects of persistency on the phase behavior of AB diblock copolymer melts.

In our present study, we employ the WLC model that covers the entire range of chain flexibility from Gaussian to semi-flexible and even rigid chains~\cite{SpakowitzPRE2005,JeffChen2016_PPS}. Hence, it is appropriate to study also low $L/a$ BCP systems.  Herein, we use the self-consistent field theory (SCFT) based on WLC model~\cite{30Matsen1996,JiangPRE2011,38Jiang2013,JiangPRL2013}  for BCPs in a thin-film setup,  with a top flat surface and a bottom corrugated substrate. We investigate the combined effects of the chain flexibility, the substrate roughness, and preference on the relative stability between the L$_{\parallel}$ and L$_{\perp}$ phases. Our findings show that the self-assembly behavior of rigid chains is \emph{distinctively different} from Gaussian chains when the chains are cast on a corrugated substrate.

This paper is organized as follows. In the next section, we introduce the self-consistent field theory (SCFT) framework and the corresponding numerical schemes for solving the WLC-SCFT equations. In section III, we present our results and discussions, and section IV contains some concluding remarks.

%%%%%%%%%%%%%%%%%%%%%%%%%%%%%%%%%%%%%%%%%%%%%%%%%
\section*{Model}
%%%%%%%%%%%%%%%%%%%%%%%%%%%%%%%%%%%%%%%%%%%%%%%%%%%%%%

We outline the self-consistent field theory (SCFT) of the continuum worm-like chain model. Consider a semi-flexible polymer melt of $n$ AB di-BCP chains confined between two surfaces. The total contour length for the entire BCP chain is $L$. Each BCP chain contains two linear blocks, $fL$ is the contour length in the A block, and $(1-f)L$ in the B block, where $f$ is the fraction of the A block. We concentrate on symmetric di-BCP, {\it i.e.}, $f=0.5$. The persistence length $\lambda$, within which the orientational correlation between monomers decays exponentially, characterizes the polymer rigidity and is assumed to be the same for the A and B blocks, $\lambda=\lambda_{\mathrm{A}}=\lambda_{\mathrm{B}}$. The effective segment length $a$ to be distinguished from the monomer size can be identified with twice the persistence $a=2\lambda$ in a worm-like chain (WLC) model, and the parameter $L/a$ becomes large for a flexible chain and small for a rod-like chain. Note that $L/a$ is equivalent to the degree of polymerization $N$ only in flexible chain limit ($L/a\gg1$)~\cite{JeffChen2016_PPS}.

In order to facilitate the numerical convergence, a masking method is used to model the surface confinement, where the impenetrable surfaces are replaced by a mask with a ``wall'' component~\cite{33Matsen1997,34BOSSe2007}. We use a local incompressibility condition $\phi_{\mathrm{A}}(\bm{\mathrm{r}})+\phi_{\mathrm{B}}(\bm{\mathrm{r}})+\phi_{\mathrm{w}}(\bm{\mathrm{r}})=1$, where $\phi_{\mathrm{A}}$ and $\phi_{\mathrm{B}}$ are the volume fractions of the A and B blocks. The third ``wall'' component fraction is $\phi_{\mathrm{W}}$, and the rigid wall is replaced by a compressible (``soft'') component characterized by an energetic penalty cost for local density deviations from the incompressibility condition. Hence, the penalty term is written as
\begin{equation}
\begin{aligned}
\zeta(\phi_{\mathrm{p}}(\bm{\mathrm{r}})+\phi_{\mathrm{w}}(\bm{\mathrm{r}})-1)^2
\end{aligned}
\end{equation}
where $\phi_{\mathrm{p}}(\bm{\mathrm{r}})=\phi_{\mathrm{A}}(\bm{\mathrm{r}})+\phi_{\mathrm{B}}(\bm{\mathrm{r}})$ is the polymer volume fraction and $\zeta$ is the energy penalty parameter.
%fig1
%%%%%%%%%%%%%%%%%%%%%%%%%%%%%%%%%%%%%%%%%%%%%%%%%%%%%%%%%%%%%%%%%%%%%%%%%%%%%%%%%%%%%%%%%%%%
\begin{figure}[h!t]
{\includegraphics[width=0.45\textwidth,draft=false]{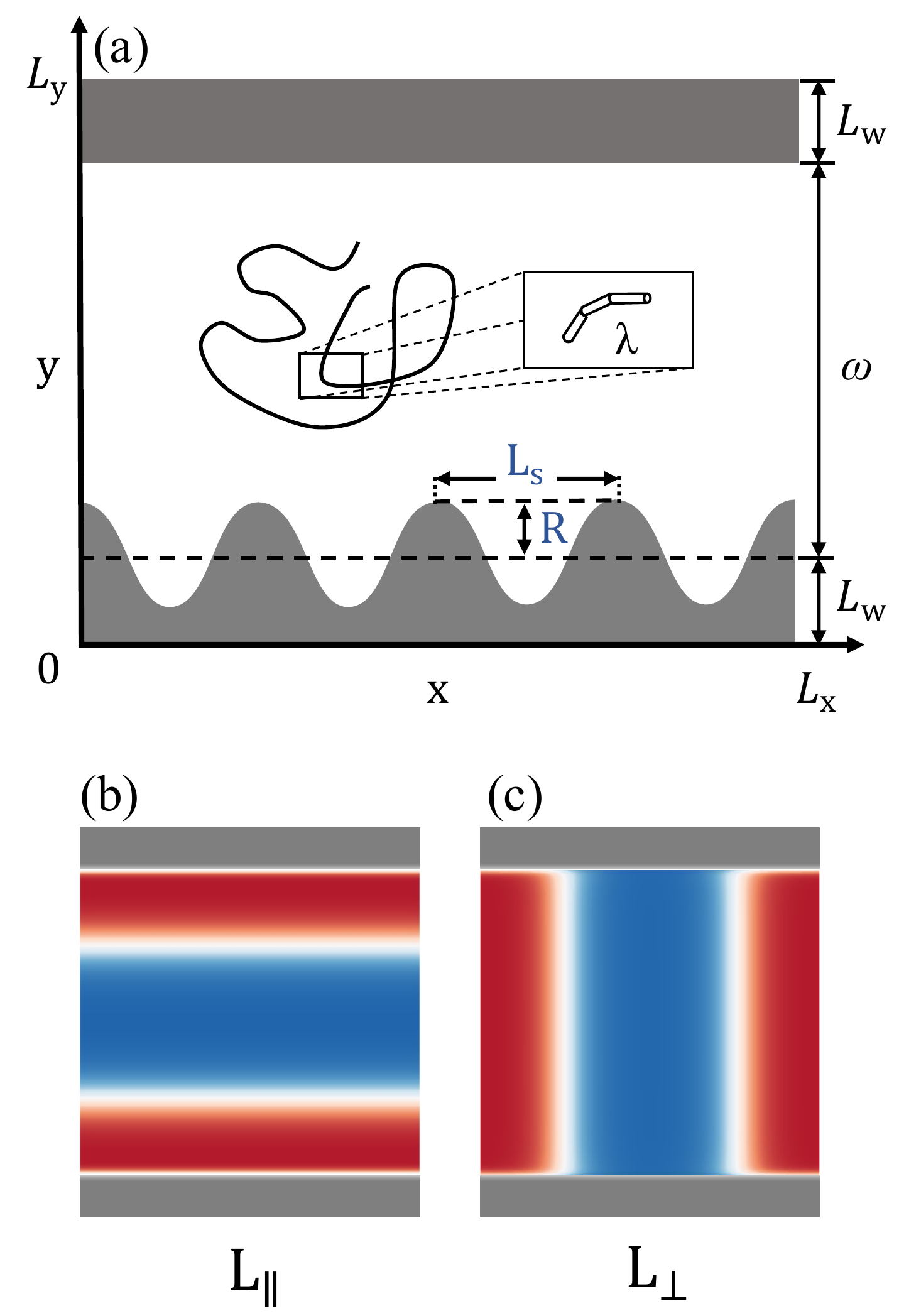}}
\caption{
\textsf{Schematic illustration of a BCP film confined between a flat top surface and a sinusoidally corrugated bottom substrate. The persistence length of the semi-flexible BCP is $\lambda$. (a) The two-dimensional simulation box has the size $L_x\times L_y$, where $\omega=L_y-2L_w$ is the averaged BCP film thickness and $L_w$ is the average substrate height. The corrugated substrate is described by a height function: $h(x)=R\cos(2\pi x/L_{\mathrm{s}})$, with periodicity $L_s$ and amplitude $R$. In (b), the lamellar orientation is parallel to the substrate (L$_{\parallel}$). In (c), the lamellar orientation is perpendicular to the substrate (L$_{\perp}$). A-rich regions are colored red and B-rich ones in blue.}}
\label{figure1}
\end{figure}
%%%%%%%%%%%%%%%%%%%%%%%%%%%%%%%%%%%%%%%%%%%%%%%%%%%%%%%%%%%%%%%%%%%%%%%%%%%%%%%%%%%%%%%%%%%%

The system is assumed to be translational invariant in the $z$-direction, so the numerical calculations are performed in an $L_{\mathrm{x}}\times L_{\mathrm{y}}$ two-dimensional box. With the above conventions, the Hamiltonian for a di-BCP film confined between two surfaces based on a WLC model is written as:
\begin{equation}
\begin{aligned}
H[W_+,W_-]&=C\int {\rm d^{3}}\bm{\mathrm{r}} \left( \frac{[W_-(\bm{\mathrm{r}})]^2}{\chi_{{\rm AB}}(L/a)}-\frac{2u(L/a)}{\chi_{{\rm AB}}(L/a)}\phi_\mathrm{w}(\bm{\mathrm{r}})W_-(\bm{\mathrm{r}})
\right.\\
&\left. +\frac{[W_+(\bm{\mathrm{r}})]^2-2\zeta (L/a)\phi_{\mathrm{p}}(\bm{\mathrm{r}})iW_+(\bm{\mathrm{r}})}{\chi_{{\rm AB}}(L/a)+2\zeta(L/a)} \right) \\
& -CV\bar{\phi}\ln Q[W_{\rm A},W_{\rm B}]
\end{aligned}
\end{equation}
where $C$ is the normalization factor, and $\phi_{\mathrm{p}}(\bm{\mathrm{r}})=\phi_{\mathrm{A}}(\bm{\mathrm{r}})+\phi_{\mathrm{B}}(\bm{\mathrm{r}})$ is the polymer volume fraction. The Flory\verb|-|Huggins interaction parameter between the A and B monomers is $\chi_{{\rm AB}}$, $u=(\chi_{{\rm wA}}-\chi_{{\rm wB}})L/a$ is the relative interaction between the substrate (w) and the A and B components, and $\chi_{{\rm wA}}$ and $\chi_{{\rm wB}}$ are the interaction parameters between the substrate and the A and B components, respectively. The polymer volume fraction averaged over the volume $V$ is $\bar{\phi}=V ^{-1}\int {\rm d^{3}}\bm{\mathrm{r}}\,\phi_{\mathrm{p}}(\bm{\mathrm{r}})$.

The quantity $Q$ is the partition function of a single copolymer chain interacting with the two conjugate fields, $W_{{\rm A}}=iW_+(\bm{\mathrm{r}})-W_-(\bm{\mathrm{r}})$ and $W_{{\rm B}}=iW_+(\bm{\mathrm{r}})+W_-(\bm{\mathrm{r}})$. It can be calculated from the integral
\begin{equation}
\begin{aligned}
Q=\frac{1}{4\pi V}\int {\rm d^{3}}\bm{\mathrm{r}}\,{\rm d^{2}}\bm{\mathrm{u}}\ q(\bm{\mathrm{r}},\bm{\mathrm{u}},s=1)
\end{aligned}
\end{equation}
where the propagator $q(\bm{\mathrm{r}},\bm{\mathrm{u}},s)$ represents the probability of finding the $s$ terminal, which is located in a spatial position specified by $\mathbf r$ and points in a direction specified by the unit vector $\mathbf u$. Here, the propagator satisfies the modified diffusion equation (MDE)~\cite{35Fredrickson}
\begin{equation}
\begin{aligned}
\frac{\partial}{\partial s}q(\bm{\mathrm{r}},\bm{\mathrm{u}},s)=\left[\frac{L}{a} \nabla ^2 _{\bm{\mathrm{u}}}-L\bm{\mathrm{u}}\cdot\nabla_{\bm{\mathrm{u}}}-W(\bm{\mathrm{r}})\right]q(\bm{\mathrm{r}},\bm{\mathrm{u}},s)
\end{aligned}
\end{equation}
The initial condition required for solving the MDE is $q(\bm{\mathrm{r}},\bm{\mathrm{u}},s=0)=1$. In addition, for BCP, $W(\bm{\mathrm{r}})=W_{{\rm A}}(\bm{\mathrm{r}})$ for $0\leq s\,\textless f$ (A block) and $W(\bm{\mathrm{r}})=W_{{\rm B}}(\bm{\mathrm{r}})$ for $f\leq s\,\textless 1$ (B block). A conjugated progapagator, $q^*(\bm{\mathrm{r}},\bm{\mathrm{u}},s)$, can be defined and starts from the final terminal ($s=1$) where $q^*(\bm{\mathrm{r}},\bm{\mathrm{u}},s=1)=1$.  The $q^*(\bm{\mathrm{r}},\bm{\mathrm{u}},s)$ satisfies a similar MDE.

In the mean-field approximation, the thermodynamic properties of the confined BCP melt can be obtained from the saddle-point configuration of the free energy in eq 2, {\it i.e.}, solving
\begin{equation}
\frac{\delta H[W_+,W_-]}{\delta (iW_+(\bm{r}))}=\frac{\delta H[W_+,W_-]}{\delta (W_-(\bm{r}))}=0
\end{equation}

The next step is to solve the MDEs for the two propagators $q(\bm{\mathrm{r}},\bm{\mathrm{u}},s)$ and $q^*(\bm{\mathrm{r}},\bm{\mathrm{u}},s)$, which depend on ${\bm{\mathrm{r}}}={\bm{\mathrm{r}}} (x,y)$, the two-dimensional orientation vector $\bm{\mathrm{u}}$, and the time-like scalar variable $s$. A detailed formulation of the numerical procedure and its implementation in SCFT modeling of BCP systems can be found elsewhere~\cite{34BOSSe2007,36Hur2009,37Takahashi2012,38Jiang2013,JiangPRL2013}.

The system setup is shown schematically in figure~\ref{figure1}a. The ``wall density'', $\phi_{\mathrm{w}}$, is fixed during the iterations. The flat top surface is characterized by a smoothly varying wall function:
\begin{equation}
\begin{aligned}
\phi_{\mathrm{w}}(y)=\frac{1}{2}+\frac{1}{2}\tanh\left[ \frac{y-L_{\mathrm{w}}-\omega}{\delta}\right]
\end{aligned}
\end{equation}
where $L_{\mathrm{w}}$ and $\omega$ are defined in figure~\ref{figure1}a, and $\delta$ is the interface width of the wall. The bottom sinusoidal substrate is described by a height function measured with respect to the average height:
\begin{equation}
\begin{aligned}
h(x)=R\cos(2\pi x/L_{\mathrm{s}})
\end{aligned}
\end{equation}
where $R$ is the amplitude of the sinusoidal corrugation wall, and $L_s$ is the corresponding periodicity. Thus, the bottom wall function is:
\begin{equation}
\begin{aligned}
\phi_{\mathrm{w}}(\bm{\mathrm{r}})=\frac{1}{2}-\frac{1}{2}\tanh \left[ \frac{y-R\cos(2\pi x/L_{\mathrm{s}})-L_{\mathrm{w}}}{\delta}\right]
\end{aligned}
\end{equation}

The SCFT formulation gives the local density of the A and B components
\begin{equation}
\begin{aligned}
\phi_{\mathrm{A}}(\bm{\mathrm{r}})=\frac{\bar{\phi}}{4\pi Q}\int {\rm d^2}\bm{\mathrm{u}}\int^f _0 {\rm ds}\,q(\bm{\mathrm{r}},\bm{\mathrm{u}},s)q^*(\bm{\mathrm{r}},\bm{\mathrm{u}},s)
\end{aligned}
\end{equation}
and
\begin{equation}
\begin{aligned}
\phi_{\mathrm{B}}(\bm{\mathrm{r}})=\frac{\bar{\phi}}{4\pi Q}\int {\rm d^2}\bm{\mathrm{u}}\int^1 _f {\rm ds}\,q(\bm{\mathrm{r}},\bm{\mathrm{u}},s)q^*(\bm{\mathrm{r}},\bm{\mathrm{u}},s)
\end{aligned}
\end{equation}
respectively. Therefore, there are two orientations of the lamellar phase with respect to the substrate, as shown schematically in figure~\ref{figure1}. The parallel orientation (L$_{\parallel}$, figure~\ref{figure1}b), and The perpendicular one (L$_{\perp}$, figure~\ref{figure1}c).

The advantage of using such a WLC model is that the chain statistics changes continuously from Gaussian to semi-flexible and even to rigid chains as the $L/a$ parameter changes from large values to small ones~\cite{JeffChen2016_PPS, 32Saito1967, 30Matsen1996,SpakowitzPRE2005,JiangPRE2011,38Jiang2013,JiangPRL2013,JeffChen2016_PPS,Rubinstein2016,Matsen2015}. In other words, the chain rigidity increases as the value of $L/a$ decreases.
In our work, we fix $\chi_{\rm AB}L/a=25$ and change $L/a$ from 100 to 2 which implies that $\chi_{\rm AB}$ changes from 0.25 to 12.5. In addition, the substrate roughness $R$ and the surface preference $u$ are changed in order to explore the relative stability of the L$_{\perp}$ and  L$_{\parallel}$ phases. It is worth noticing that small values of the parameter $L/a$ are practically meaningful for real polymer systems. According to experimental measurements, the value of $L/a$ roughly ranges from $1$ to $10$ for conjugated block copolymers~\cite{LEE2015,Gettinger1994,McCulloch2013}, block bottlebrush copolymers~\cite{Fei2019} and liquid crystalline polymers~\cite{Feng2016}.

%%%%%%%%%%%%%%%%%%%%%%%%%%%%%%%%%%%%%%%%%%%%%%%%%%
\section*{Results And Discussion}
%%%%%%%%%%%%%%%%%%%%%%%%%%%%%%%%%%%%%%%%%%%%%%%%%%

\subsection* {Varying the Chain Flexibility (L/a) for Flat Bounding Walls}

We first discuss how varying the chain flexibility affects the relative stability of the parallel (L$_{\parallel}$) and the perpendicular (L$_{\perp}$) orientations, for BCP thin films, confined between two {\it flat and neutral} walls.  The BCPs form a lamellar phase in the bulk with a periodicity $d_0$\,=\,$d_0$ that is a function of the chain flexibility parameter, $L/a$. For flexible polymers ($L/a$\,=\,$100$), the lamellar periodicity $d_0/a$\,=\,$17.2$,  is consistent with previous calculations based on the Gaussian model, $d_0$\,=\,$4.25 R_g$ ($R_g$\,=\,$a\sqrt{N/6}$ is the chain radius of gyration). For chain flexibilities $L/a$\,=\,$20, 10, 5, 3,2$ and keeping $a$ fixed, the obtained periodicities are, respectively, $d_0/a$\,=\,$7.48, 5.07, 3.27, 2.25, 1.62$. The calculation box size for BCPs with different $L/a$ is chosen according to their corresponding $d_0$. In the following, the lateral size of the calculation box is fixed, $L_x$\,=\,$d_0$, for all calculations.

\begin{figure*}[h!t]
{\includegraphics[width=0.95\textwidth,draft=false]{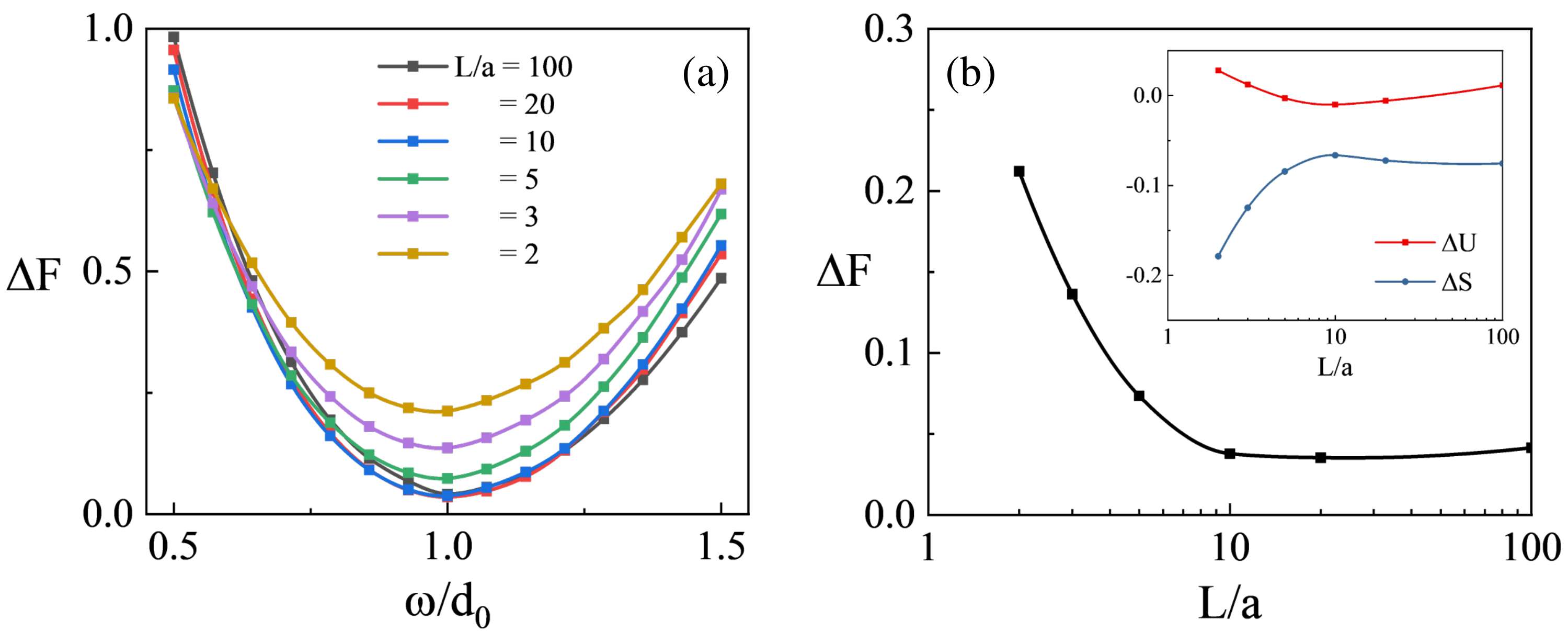}}
\caption{
\textsf{(a) The free energy difference $\Delta F=F_{\parallel}-F_{\perp}$ between the L$_{\parallel}$ and L$_{\perp}$ orientations of BCP lamellar, in units of $nk_BT$, as a function of the film thickness, $\omega/d_0$, where $n$ is the number of chains in the system, $k_BT$ is the thermal energy, and $d_0$ is the  lamellar periodicity. The parallel free energy, $F_{\parallel}$, is calculated for one parallel lamella confined between two flat and neutral surfaces, with $u=R=0$. The perpendicular free energy, $F_{\perp}$, corresponds to a perfect perpendicular lamellar phase. (b) Free energy ($\Delta F$), enthalpy ($\Delta U$) and entropy ($\Delta S$) difference (inset) between the L$_{\parallel}$ and L$_{\perp}$ lamellar orientations, where $\Delta U=U_{\parallel}-U_{\perp}$, in units of $nk_BT$, and $\Delta S=S_{\parallel}-S_{\perp}$, in units of $nk_B$,  as a function of the chain flexibility $L/a$. The other parameters are $L_x=\omega=d_0$ and $u=R=0$.}}
\label{figure2}
\end{figure*}

When the BCP film thickness, $\omega$, deviates from an integer multiple of $d_0$, the polymer chains in the L$_{\parallel}$ phase will be stretched or compressed. This is seen in figure~\ref{figure2}a from the free-energy difference $\Delta F $\,=\,$ F_{\parallel}-F_{\perp}$ between the parallel free-energy, $ F_{\parallel}$ and the perpendicular one, $ F_{\perp}$. Two significant features can be seen in figure~\ref{figure2}a. (i) The free-energy difference $\Delta F$ has a local minimum when the film thickness $\omega$ equals the natural periodicity $d_0$; and, (ii) the free energy $F_{\perp}$ is strictly lower than $F_{\parallel}$, and the value of the local minimum in $\Delta F$ varies for different $L/a$. The first feature can be understood because the residual elastic strains due to confinement will be suppressed when the film thickness equals an integer multiple of the natural periodicity $d_0$.

%fig2
%%%%%%%%%%%%%%%%%%%%%%%%%%%%%%%%%%%%%%%%%%%%%%%%%%%%%%%%%%%%%%%%%%%%%%%%%%%%%%%%%%%%%%%%%%%%
%%%%%%%%%%%%%%%%%%%%%%%%%%%%%%%%%%%%%%%%%%%%%%%%%%%%%%%%%%%%%%%%%%%%%%%%%%%%%%%%%%%%%%%%%%%%

To understand the second feature, we show the dependence of the $\Delta F$ local minimum on $L/a$ in figure~\ref{figure2}b. For rigid or semi-flexible chains ($L/a$\,=\,$2, 3$ and $5$), the local minimum of $\Delta F$ at film thickness $\omega/d_0$\,=\,$1$ increases as $L/a$ decreases. When the polymer is fairly flexible ($L/a$\,=\,$10, 20,$ and $100$), the local minimum of $\Delta F$ remains nearly unchanged. The inset in figure~\ref{figure2}b shows the dependence on $L/a$ of the entropy difference $\Delta S$ and enthalpy difference $\Delta U$ between the L$_{\parallel}$ and L$_{\perp}$ phases, respectively. 

The numerical results indicate that the change of $\Delta F=\Delta U-T\Delta S$ is mainly caused by $\Delta S$ because $\Delta U$ remains nearly zero as $L/a$ changes. In the calculations, the dimensionless entropy is
\begin{equation}
\begin{aligned}
\frac{S}{nk_B}=\frac{1}{\bar{\phi}V}\int {\rm d^3}\bm{\mathrm{r}}(\phi_{\mathrm{A}}W_{\mathrm{A}}+\phi_{\mathrm{B}}W_{\mathrm{B}})\,+\,\ln Q,
\end{aligned}
\end{equation}
and the dimensionless enthalpy is
\begin{equation}
\begin{aligned}
\frac{U}{nk_BT}=\frac{1}{\bar{\phi}V}\int {\rm d^3}\bm{\mathrm{r}} \chi_{\mathrm{AB}}(L/a)\phi_{\mathrm{A}}\phi_{\mathrm{B}}.
\end{aligned}
\end{equation}

Figure~\ref{figure2}b indicates that the L$_{\perp}$ phase is always more stable than L$_{\parallel}$. Moreover, the stability of L$_{\perp}$ increases as $L/a$ decreases. These results are consistent with previous studies,~\cite{39Pickett, 40Pickett2} where it was shown that for two flat and neutral surfaces, the perpendicular orientation is favored over the parallel orientation because of the entropic confinement effect. This is the so-called ``\emph{nematic effect}'' where a hard wall limits the chain conformations and facilitates the chain stretching along the substrate~\cite{JeffChenMM2006,JeffChen2016_PPS}. Moreover, Pickett {\it {et al.}}~\cite{40Pickett2} showed that the energy difference between the parallel and perpendicular orientations scales as $N^{-2/3}$. Therefore, $\Delta F$ increases as $N$ decreases, which is also consistent with our full numerical calculations.

%fig3
%%%%%%%%%%%%%%%%%%%%%%%%%%%%%%%%%%%%%%%%%%%%%%%%%%%%%%%%%%%%%%%%%%%%%%%%%%%%%%%%%%%%%%%%%%%%
\begin{figure}[h!t]
{\includegraphics[width=0.45\textwidth,draft=false]{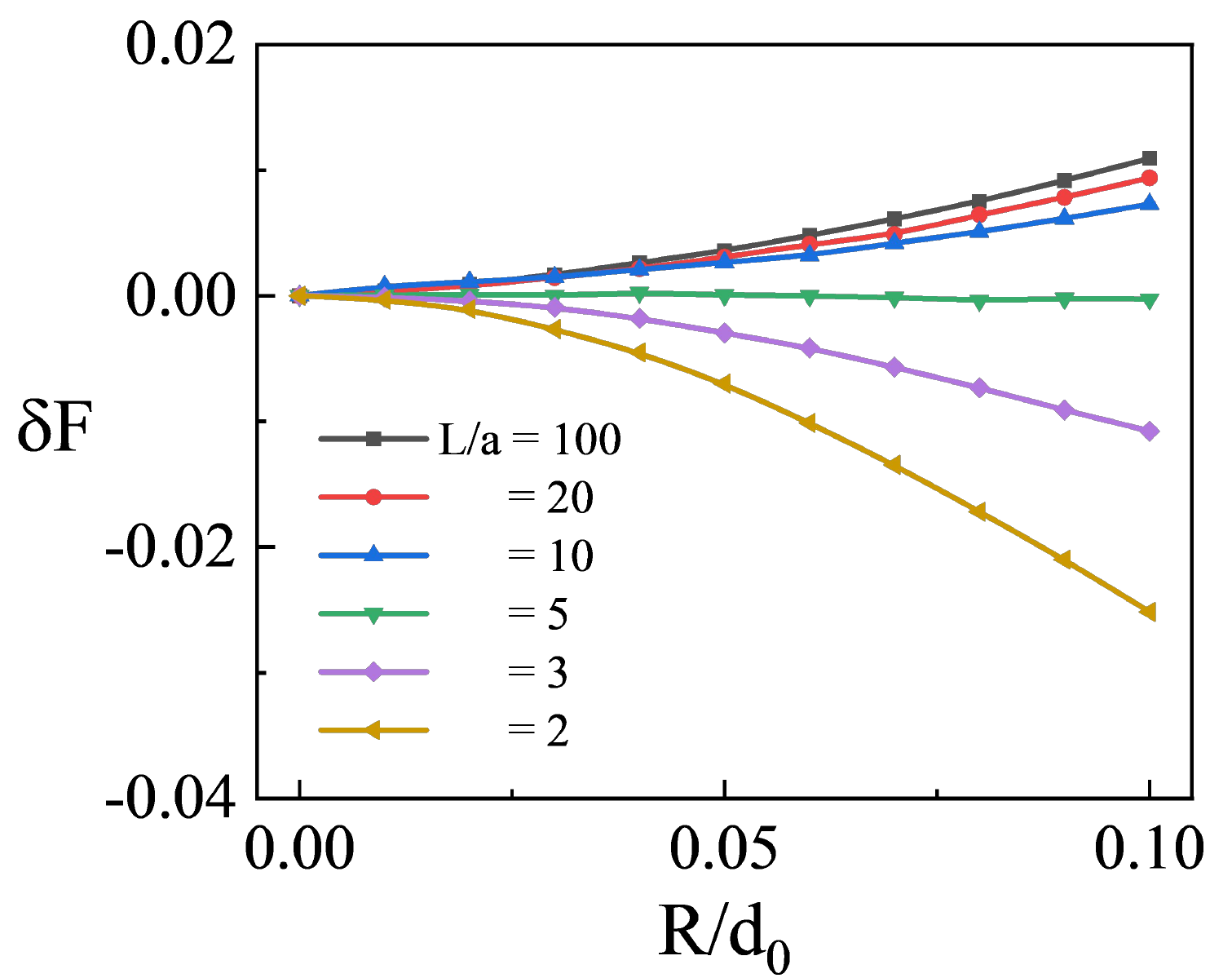}}
\caption{
\textsf{Dependence of $\delta F=\Delta F(R)-\Delta F$($R$\,=\,$0$) on the amplitude $R$ for different chain flexibility $L/a$, where the reference $\Delta F$($R$\,=\,$0$) is calculated for a flat ($R$\,=\,$0$) and neutral ($u$\,=\,$0$) substrate. Other parameters are $L_s=d_0$, $\omega=L_x=d_0$.}}
\label{figure3}
\end{figure}
%%%%%%%%%%%%%%%%%%%%%%%%%%%%%%%%%%%%%%%%%%%%%%%%%%%%%%%%%%%%%%%%%%%%%%%%%%%%%%%%%%%%%%%%%%%%

\subsection*{Substrate Roughness Effect}

We investigate the combined effects of the wall roughness amplitude $R$ and the BCP chain flexibility $L/a$ on the relative stability between the L$_{\parallel}$ and L$_{\perp}$ phases. We fix the lateral size of the calculation box ($L_x$) and the corrugation periodicity ($L_s$), $L_x$\,=\,$L_s$\,=\,$d_0$, where $d_0$\,=\,$d_0$ is a function of the $L/a$. We change the rescaled amplitude $R/d_0$ to obtain substrates with various roughness. It is clear that the order of magnitude of $\Delta F=F_{\parallel}-F_{\perp}$ is quite different for BCPs with different chain flexibility. Therefore, in order to compare the effect of $R$ on the relative stability between the two lamellar phases for BCPs with different $L/a$, we analyze $\Delta F$ for each $L/a$ by subtracting its corresponding free-energy difference for a flat substrate and the same $L/a$, $\delta F(R,L/a)$\,=\,$\Delta F(R,L/a)-\Delta F$($R$\,=\,$0,L/a$), indicating that L$_{\perp}$ becomes more stable than L$_{\parallel}$ as $\delta F$ increases. It is worth noticing that for neutral surfaces, we obtain $\Delta F(R, L/a)>0$ in all calculations. This result means that although the strength of relative stability between the two phases changes with $R$, the L$_{\perp}$ phase is always more stable than the L$_{\parallel}$ phase.  

%In these cases, although the strength of the relative stability decreases, the L$_{\perp}$ is still more stable than the L$_{\parallel}$ because $\Delta F$ remains being positive in all calculations.

One of our main results is shown in figure~\ref{figure3} where we plot $\delta F$ as a function of the roughness amplitude $R$ for $L/a=2,3,5,10,20$, and $100$. With the increase of the roughness amplitude $R$, $\delta F$ manifests diverse variation tendencies for different chain flexibility, $L/a$. In the case of Gaussian chains ($L/a$\,$\ge$\,$10$), $\delta F$ increases as $R$ increases. In the case of semi-flexible polymer ($L/a$\,=\,$5$), $\delta F$ remains  nearly unaffected as $R$ increases. However, for rigid chains, $L/a$\,=\,$2$ and $3$, $\delta F$ decreases with the increase of $R$.

%fig4
%%%%%%%%%%%%%%%%%%%%%%%%%%%%%%%%%%%%%%%%%%%%%%%%%%%%%%%%%%%%%%%%%%%%%%%%%%%%%%%%%%%%%%%%%%%%
\begin{figure*}[h!t]
{\includegraphics[width=1\textwidth,draft=false]{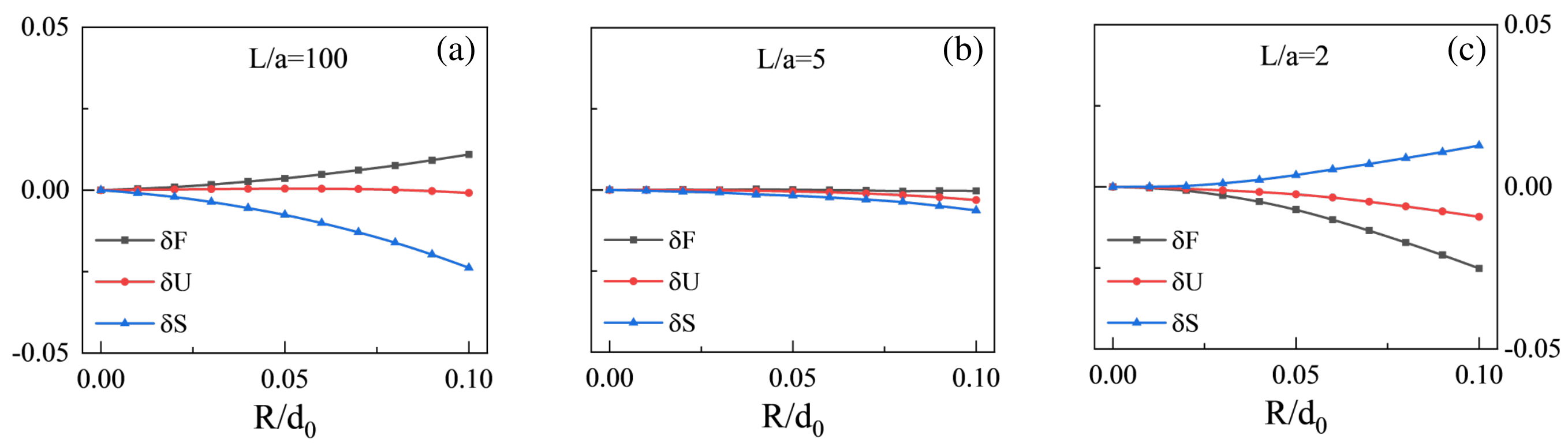}}
\caption{
\textsf{Dependence of $\delta F=\Delta F(R)-\Delta F$($R$\,=\,$0$), $\delta U=\Delta U(R)-\Delta U$($R$\,=\,$0$), and $\delta S=\Delta S(R)-\Delta S$($R$\,=\,$0$) on the amplitude $R$ for various chain flexibilities (a) $L/a=100$, (b) $L/a=5$, and (c) $L/a=2$. $\Delta F$, $\Delta U$ and $\Delta S$ are the same as in figure~\ref{figure2}.}}
\label{figure4}
\end{figure*}
%%%%%%%%%%%%%%%%%%%%%%%%%%%%%%%%%%%%%%%%%%%%%%%%%%%%%%%%%%%%%%%%%%%%%%%%%%%%%%%%%%%%%%%%%%%%

To understand these different trends, we analyze separately the enthalpy and entropy contributions to the free energy. Figure~\ref{figure4} shows the dependence of $\delta F$, $\delta U$ and $\delta S$ on $R/d_0$ for $L/a$\,=\,$2,  5$ and $100$, separately. For each of these three quantities, we subtract their corresponding values for a flat substrate ($R$\,=\,$0$), as is defined in the caption of Figure~\ref{figure4}, {\it i.e.}, $\delta F=\Delta F(R)-\Delta F$($R$\,=\,$0$).

%fig5
%%%%%%%%%%%%%%%%%%%%%%%%%%%%%%%%%%%%%%%%%%%%%%%%%%%%%%%%%%%%%%%%%%%%%%%%%%%%%%%%%%%%%%%%%%%%
\begin{figure*}[h!t]
{\includegraphics[width=0.7\textwidth,draft=false]{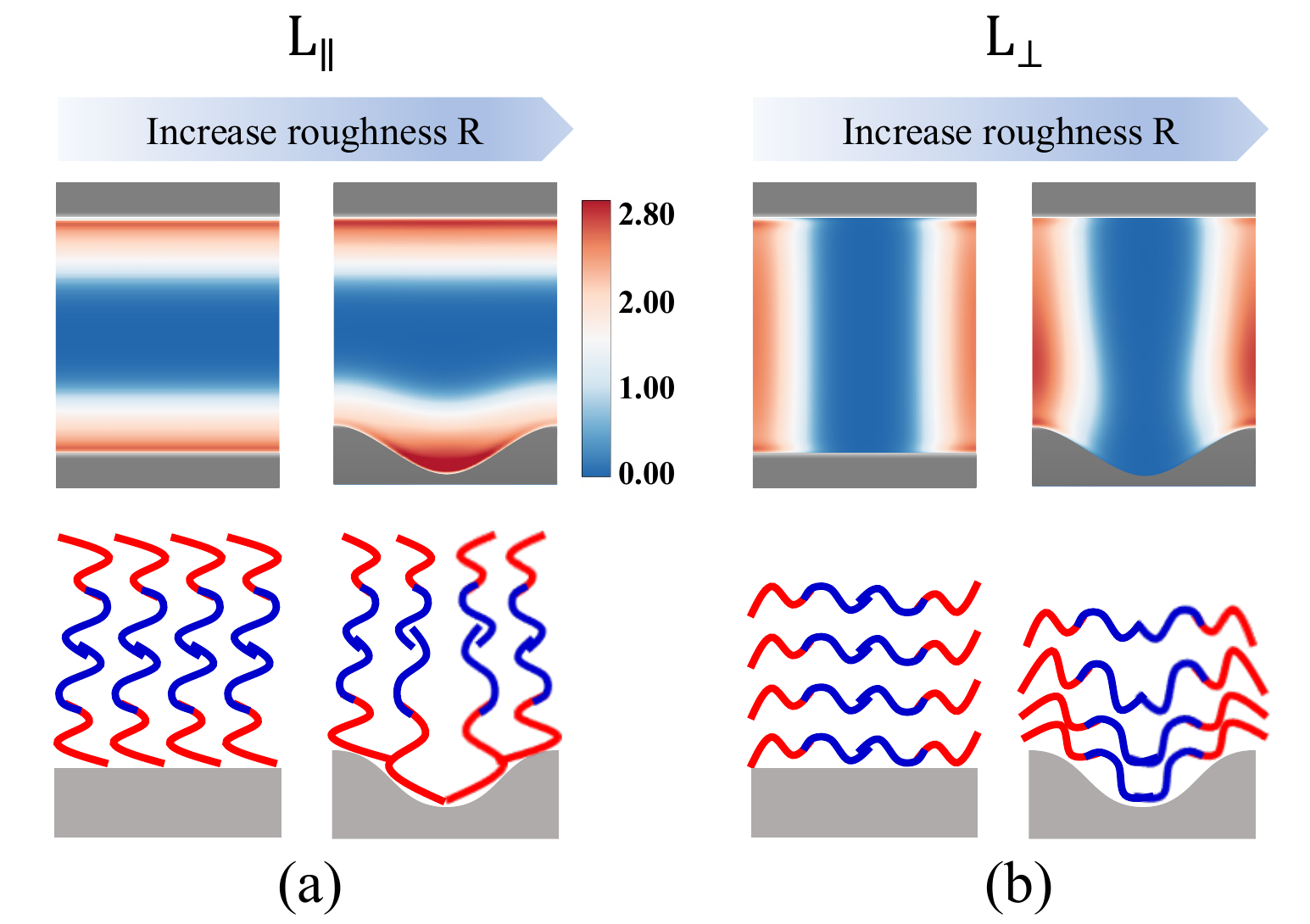}}
\caption{
\textsf{The distribution of the chain end location (top) and schematic illustration of the chains (bottom) for $L/a$\,=\,$100$ chains. (a) The L$_{\parallel}$ lamellar phase, and (b) the L$_{\perp}$ lamellar phase are shown separately for flat ($R$\,=\,$0$) and corrugated ($R$\,=\,$0.1d_0$) substrates.}}
\label{figure5}
\end{figure*}
%%%%%%%%%%%%%%%%%%%%%%%%%%%%%%%%%%%%%%%%%%%%%%%%%%%%%%%%%%%%%%%%%%%%%%%%%%%%%%%%%%%%%%%%%%%%

Figure~\ref{figure4}a shows the behavior for the Gaussian chain with $L/a$\,=\,$100$. With the increase of the rescaled amplitude $R/d_0$, the entropy difference $\delta S$ (blue line) decreases, while the enthalpy difference $\delta U$ (red line) remains nearly unchanged. This can be understood in more detail by analyzing the chain configuration in the thin film geometry. Figure~\ref{figure5}a shows the distribution of the chain ends for the L$_{\parallel}$ phase on both flat and corrugated substrates. Our calculations show that the chain ends are evenly distributed in the entire thin film for a flat substrate. However, for corrugated substrates, the chain ends slide down into the valleys to accommodate the surface corrugations. In other words, the polymer chains are compressed due to the surface corrugation, indicating that increasing $R$ leads to a decrease in chain configuration entropy of the L$_{\parallel}$ phase. On the other hand, figure~\ref{figure5}b shows that the chain end distribution for the L$_{\perp}$ phase on a corrugated substrate is nearly unchanged as compared to a flat substrate. The reason for this difference in behavior is that the polymer chains are flexible and lie down in the L$_{\perp}$ phase, as they can adjust easily to the surface corrugation. The outcome of these two different chain responses is that the decrease of $S_{\parallel}$ is larger than $S_{\perp}$, leading to a decrease in $\delta S$ as the surface roughness amplitude $R$ increases.

%hence,  $S_{\parallel}$ decreases markedly. the polymer chains in the L$_{\perp}$ phase are lying down and are less compressed due to the surface corrugations.  

%At the same time, $\delta U$ is nearly unchanged, and $\delta F$ increases as $R$ increases, indicating that the perpendicular lamellar phase L$_{\perp}$ becomes more stable than the parallel L$_{\parallel}$ one.

\begin{figure*}[h!t]
{\includegraphics[width=0.7\textwidth,draft=false]{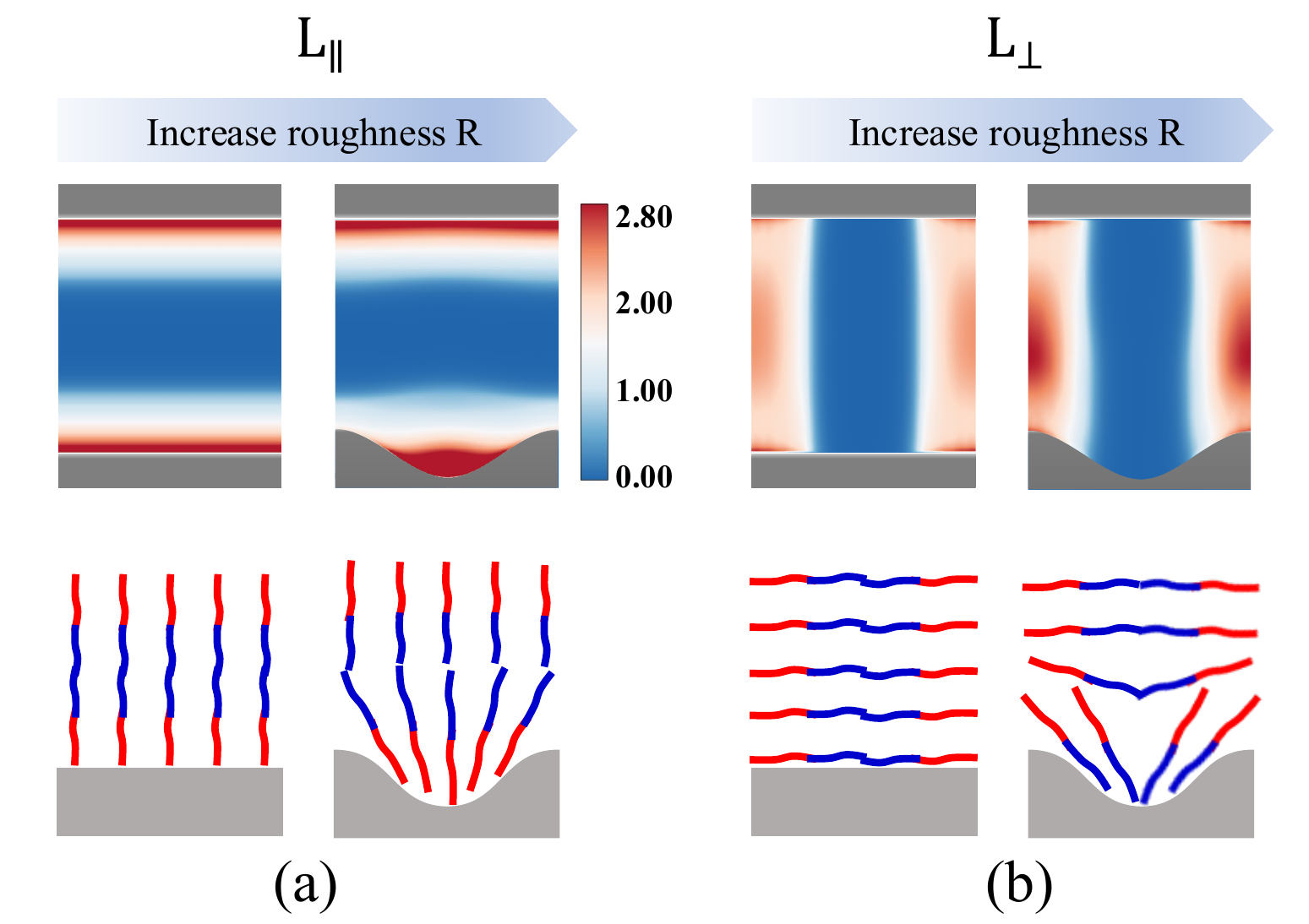}}
\caption{
\textsf{ The distribution of the chain-end location (top) and schematic illustration of the chains (bottom) for $L/a$\,=\,$2$ chains. (a) The L$_{\parallel}$ lamellar phase, and (b) the L$_{\perp}$ lamellar phase are shown separately for flat ($R$\,=\,$0$) and corrugated ($R$\,=\,$0.1d_0$) substrates. }}
\label{figure6}
\end{figure*}

Figure~\ref{figure4}c shows the case of the rigid chain ($L/a$\,=\,$2$). As $R/d_0$ increases, $\delta F$ decreases and $\delta S$ increases, which is \emph{completely opposite} to the behavior of Gaussian chains ($L/a$\,=\,$100$). Such behavior is closely related to the chain configurations for small $L/a$. Figure~\ref{figure6} presents the calculated distribution of the chain ends for $L/a$\,=\,$2$. Figure~\ref{figure6}a shows that for the L$_{\parallel}$ phase, the chain ends slide down into the valleys when BCPs are cast on top of a corrugated substrate. However, the polymer chains retain their standing-up configuration when they are cast on a flat substrate. By contrast, for the L$_{\perp}$ orientation, figure~\ref{figure6}b shows that the chain ends of the A-block concentrate in the middle of the thin film, indicating that the entire BCP chains stand up because the chains behave as rigid rods. As a consequence, increasing $R$ enhances the chain configuration distortion of the L$_{\perp}$ phase because the chains lie parallel to the flat surface, leading to a markedly decrease in the $S_{\perp}$ entropy. Thus, the decrease of $S_{\parallel}$ is smaller than in $S_{\perp}$, and $\delta S=S_{\parallel}-S_{\perp}$ increases with $R$. Consequently, $\delta F$ decreases, indicating that the relative stability of L$_{\perp}$ decreases.

%fig6
%%%%%%%%%%%%%%%%%%%%%%%%%%%%%%%%%%%%%%%%%%%%%%%%%%%%%%%%%%%%%%%%%%%%%%%%%%%%%%%%%%%%%%%%%%%%
%%%%%%%%%%%%%%%%%%%%%%%%%%%%%%%%%%%%%%%%%%%%%%%%%%%%%%%%%%%%%%%%%%%%%%%%%%%%%%%%%%%%%%%%%%%%

%By contrast, for the L$_{\parallel}$ phase, $S_{\parallel}$ decreases slightly because the polymer chains that are initially perpendicular to the flat surface suffer less distortion on the rough substrate, as shown in Fig.~\ref{figure6}a. 

BCP chains of $L/a$\,=\,$5$, shown in figure~\ref{figure4}b, manifest an intermediate behavior. As two different tendencies compete with each other, $\delta S$ and $\delta F$ nearly do not vary with $R$. Thus, the relative stability of the two phases does not change.

The self-assembly behavior of BCP thin films made of flexible chains (large $L/a$) is consistent with previous SCFT calculations~\cite{20Man2015}, where it was found that rough substrates can enhance the stability of the L$_{\perp}$ phase. However, in the present study, we show that the response to the substrate roughness of rigid BCPs having a small $L/a$ is opposite to that of long Gaussian chains. In the latter case, large surface roughness destabilizes the stability of the L$_{\perp}$ phase, and there is a tendency for the L$_{\perp}$ phase to undergo a transition into an L$_{\parallel}$ one.

We note that the accurate quantitive comparison between the different behavior of rigid and Gaussian chains is the main finding of our study. Such findings have not been previously reported either in analytical works or in numerical simulations. These predictions should be verified in future experiments.

%%%%%%%%%%%%%%%%%%%%%%%%%%%%%%%%%%%%%%%%%%%%%%%%%%%%%%%%%%%%%%%%%%%%%%%%%%%%%%%%%%%%%%%%%%%%

%%%%%%%%%%%%%%%%%%%%%%%%%%%%%%%%%%%%%%%%%%%%%%
\subsection*{The dependence of the L$_{\perp}$-to-L$_{\parallel}$ phase transition on the Substrate preference}
%%%%%%%%%%%%%%%%%%%%%%%%%%%%%%%%%%%%%%%%%%%%%
As mentioned above, the substrate preference $u$ is an important parameter to consider in tuning the L$_{\perp}$-to-L$_{\parallel}$ phase transition, especially for BCP with high $\chi$ and low $N$, whose constituent blocks usually have considerably different surface energies. We keep the top surface neutral and flat while fixing the corrugation periodicity, $L_x$\,=\,$L_s$\,=\,$d_0$, of the bottom substrate. We then investigate the dependence of $u^{*}$, the critical value of the substrate preference, on the amplitude value $R$, where $u^{*}$ is the critical value leading to the L$_{\perp}$-to-L$_{\parallel}$ phase transition on the amplitude value $R$.

%fig7
%%%%%%%%%%%%%%%%%%%%%%%%%%%%%%%%%%%%%%%%%%%%%%%%%%%%%%%%%%%%%%%%%%%%%%%%%%%%%%%%%%%%%%%%%%%%
\begin{figure}[h!t]
{\includegraphics[width=1\textwidth,draft=false]{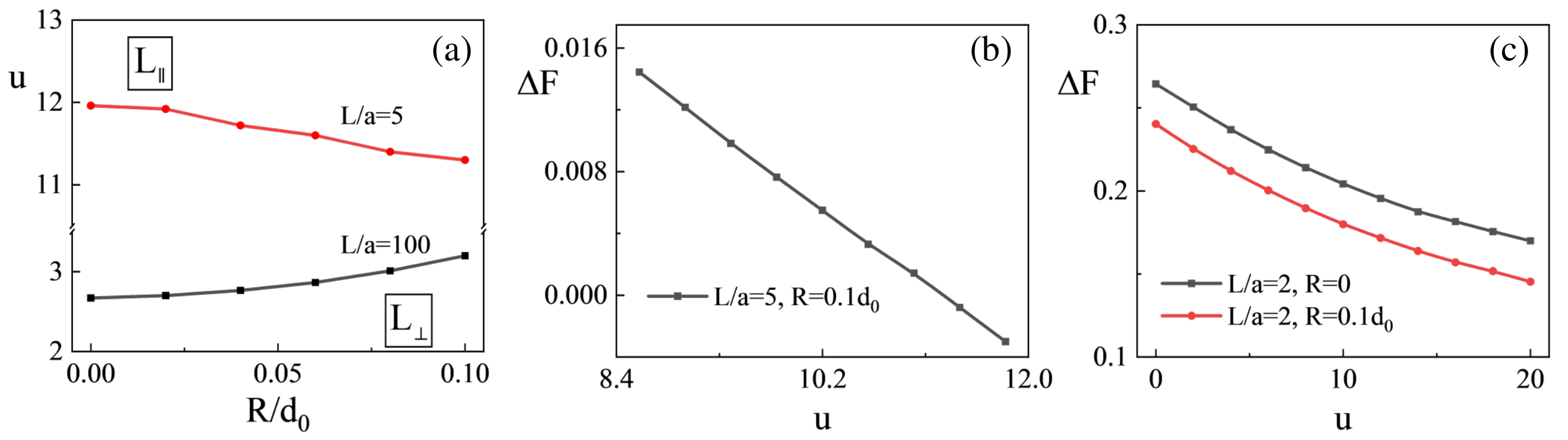}}
\caption{
\textsf{(a) The L$_{\perp}$-to-L$_{\parallel}$ phase diagram in the ($u$, $R$) plane, for two chain flexibilities: $L/a$\,=\,$100$ (black line), and $L/a=5$ (red line). (b) The dependence of $\Delta F$\,=\,$F_{\parallel}-F_{\perp}$ on the substrate preference $u$ for $L/a$\,=\,$5$ for roughness amplitude $R$\,=\,$0.1d_0$. (c) The dependence of $\Delta F$\,=\,$F_{\parallel}-F_{\perp}$ on the substrate preference $u$ for $L/a$\,=\,$2$. The flat surface ($R$\,=\,$0$) is denoted by a black line and the corrugate one ($R$\,=\,$0.1d_0$)  by a red line.}}
\label{figure7}
\end{figure}
%%%%%%%%%%%%%%%%%%%%%%%%%%%%%%%%%%%%%%%%%%%%%%%%%%%%%%%%%%%%%%%%%%%%%%%%%%%%%%%%%%%%%%%%%%%%

Figure~\ref{figure7}a shows the phase transition (L$_{\perp}$-to-L$_{\parallel}$) between the perpendicular and parallel orientations in terms of the rescaled amplitude, $R/d_0$, and the substrate preference, $u$. Any combination of $u$ and $R/d_0$ values above the plotted transition line induces an L$_{\perp}$-to-L$_{\parallel}$ phase transition. We show an example for $L/a$\,=\,$5$ and $R/d_0$\,=\,$0.1$ in figure~\ref{figure7}b. The value of $\Delta F$ changes from being positive to negative, indicating that a phase transition from L$_{\perp}$ to L$_{\parallel}$ occurs with the increase of $u$.

In addition, one can see in figure~\ref{figure7}a that Gaussian chains having large $L/a$\,=\,$100$, the critical value, $u^{*}$, corresponding to the L$_{\perp}$-to-L$_{\parallel}$ phase transition, increases as a function of $R/d_0$. For Gaussian polymers, the L$_{\perp}$ phase becomes more stable as the substrate roughness increases. Therefore, larger surface preference is needed to induce the L$_{\perp}$-to-L$_{\parallel}$ phase transition for larger substrate roughness. However, in the case of semi-flexible chains ($L/a$\,=\,$5$), there is a slight negative correlation that $u^{*}$ decreases as $R/d_0$ increases. The reason is that the L$_{\perp}$ becomes less stable as $R$ increases for semi-flexible and rigid polymers, resulting in a decrease of $u^{*}$. For rigid chains ($L/a=2$), we do not find any meaningful value of $u$ that can induce the L$_{\perp}$-to-L$_{\parallel}$ phase transition. Furthermore, large $u$ weakens the relative stability of the L$_{\perp}$ as compared with L$_{\parallel}$. As shown in figure~\ref{figure7}c,  $\Delta F$ of $L/a$\,=\,$2$ case decreases as $u$ increases. This result means that increasing $u$ and $R$ are both favoring the L$_{\parallel}$ phase for rigid polymers, but the effect is not strong enough to stabilize the L$_{\parallel}$ phase as compared with the L$_\perp$ phase.

%%%%%%%%%%%%%%%%%%%%%%%%%%%%%%%%%%%%%%%%%%%%%%%%%%
\section*{Conclusions}
%%%%%%%%%%%%%%%%%%%%%%%%%%%%%%%%%%%%%%%%%%%%%%%%%%

\begin{figure}[h!t]
{\includegraphics[width=0.7\textwidth,draft=false]{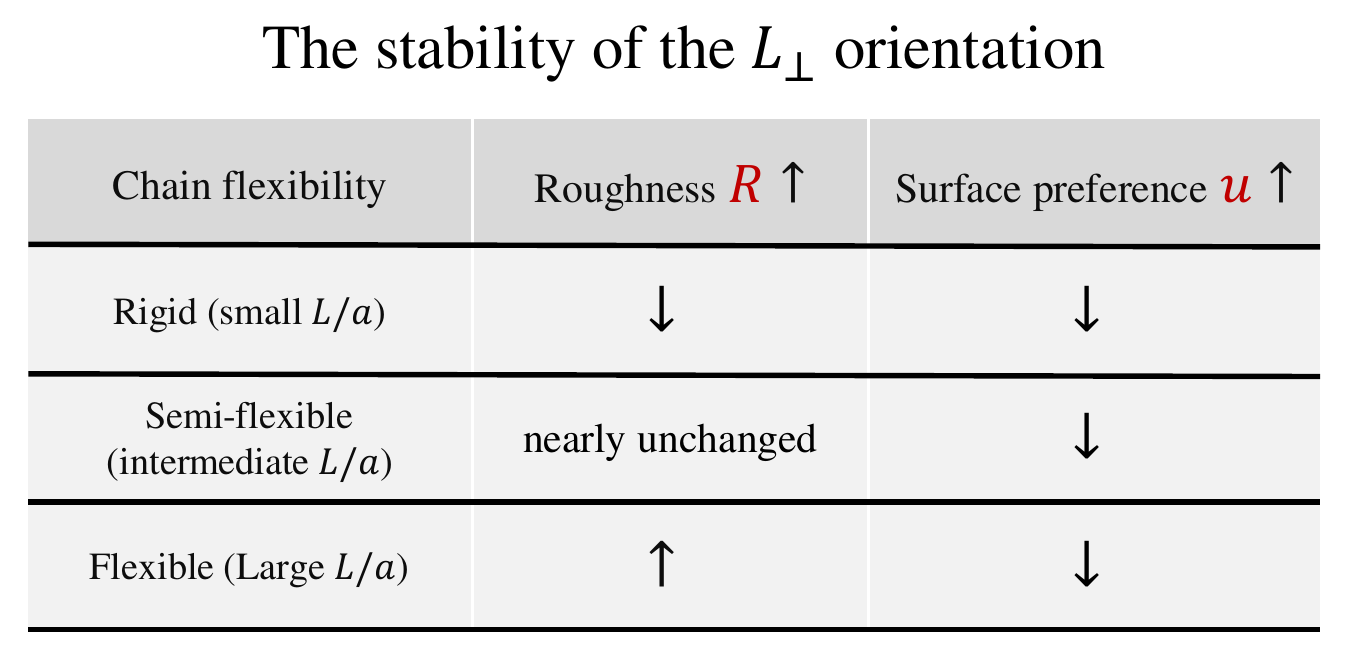}}
\caption{
\textsf{The combined effect of the chain flexibility $L/a$, substrate roughness $R$, and surface preference $u$ on the stability of the L$_{\perp}$ orientation.}}
\label{figure8}
\end{figure}

We explore the self-assembly of symmetric AB diblock copolymers (BCP) confined in a thin-film geometry. Using SCFT which is based on a continuum WLC model, we focus on the influence of the substrate structure and chemistry on the conformations of polymer chains with various flexibilities. We systematically studied the combined effect of the polymer chain flexibility parameterized by $L/a$, substrate roughness amplitude $R$, and the surface preference $u$ on the relative stability between the parallel (L$_{\parallel}$) and perpendicular (L$_{\perp}$) lamellar phases. The effects of these parameters on the relative stability of the L$_{\perp}$ phase are presented in figure~\ref{figure8}.

%fig8
%%%%%%%%%%%%%%%%%%%%%%%%%%%%%%%%%%%%%%%%%%%%%%%%%%%%%%%%%%%%%%%%%%%%%%%%%%%%%%%%%%%%%%%%%%%%
%%%%%%%%%%%%%%%%%%%%%%%%%%%%%%%%%%%%%%%%%%%%%%%%%%%%%%%%%%%%%%%%%%%%%%%%%%%%%%%%%%%%%%%%%%%%

Our results show that for BCP films confined between two flat and neutral substrates, the lamellae domains tend to orient in the L$_{\perp}$ direction, and decreasing the polymer chain flexibility will intensify this effect. However, increasing the substrate roughness $R$ has distinctively different effects on polymers of different flexibility. While for Gaussian polymer chains (large $L/a$), the stability of the L$_{\perp}$ phase is enhanced, the stability of the L$_{\perp}$ phase is weakened for rigid polymer chains (small $L/a$). We further show that as $R$, the substrate roughness, increases, the critical value of the substrate preference, $u^*$, corresponding to the L$_{\perp}$-to-L$_{\parallel}$ orientational transition decreases for rigid chains but increases for Gaussian chains. Such an opposite effect of substrate roughness for rigid chain polymers as compared to flexible ones was not observed in previous works and is one of our main results.

The origin of the distinctive behavior for Gaussian and semi-flexible (or rigid) BCP chains is obtained by analyzing the entropy and enthalpy contributions to the free energy. For Gaussian polymer chains, the roughness of the substrate influences more the conformational entropy of BCP chains in the L$_{\parallel}$ as compared with the L$_{\perp}$ due to the chain flexibility. Thus, the L$_{\perp}$ stability is enhanced. Nevertheless, for rigid polymer chains, large roughness amplitude, $R$, affects more the chain conformational entropy in the L$_{\perp}$ than in the L$_{\parallel}$ due to the chain rigidity. Therefore, the effects of surface roughness on polymer conformational entropy weaken the L$_{\perp}$ stability for rigid BCPs.

In conclusion, our study systematically manifests that rigid BCP chains (or short chains) are more likely to form an L$_{\perp}$ phase in thin films, as is desirable in nano-lithography applications that are used to generate sub-10nm patterns. Using a corrugated substrate to induce a perpendicular orientation of the BCP film is useful for Gaussian BCP chains but will not work for rigid chains. We hope that our results may serve as a useful guide for future modeling experiments and applications.

%%%%%%%%%%%%%%%%%%%%%%%%%%%%%%%%%%%%%%%%%%%%%%%%%%

\bigskip
%%%%%%%%%%%%%%%%%%%%%%%%%%%%%%%%
{\bf Acknowledgement.}~~
%%%%%%%%%%%%%%%%%%%%%%%%%%%%%%%%
This work was supported in part by the NSFC-ISF Research Program, jointly funded by the National Natural Science Foundation of China (NSFC) under grant No.~21961142020 and the Israel Science Foundation (ISF) under grant No.~3396/19, NSFC grants No.~21822302 and 22073004 and  ISF grant No. 213/19, the Fundamental Research Funds for the Central University under grant No.~YWF-22-K-101. We also acknowledge the support of the High-Performance Computing Center of Beihang University.
%\newpage

%%%%%%%%%%%%%%%%%%%%%%%%%%%%%%%%%%%%%%%%%

\clearpage
\vskip 0.5truecm
\centerline{for Table of Contents use only}
\centerline{\bf The Chain Flexibility Effects on the Self-assembly of Diblock Copolymer in Thin Film}
\centerline{\it Mingyang Chen, Yuguo Chen, Yanyan Zhu, Ying Jiang$^*$, David Andelman$^*$, and Xingkun Man$^*$}

\begin{figure}[h]
{\includegraphics[width=0.7\textwidth,draft=false]{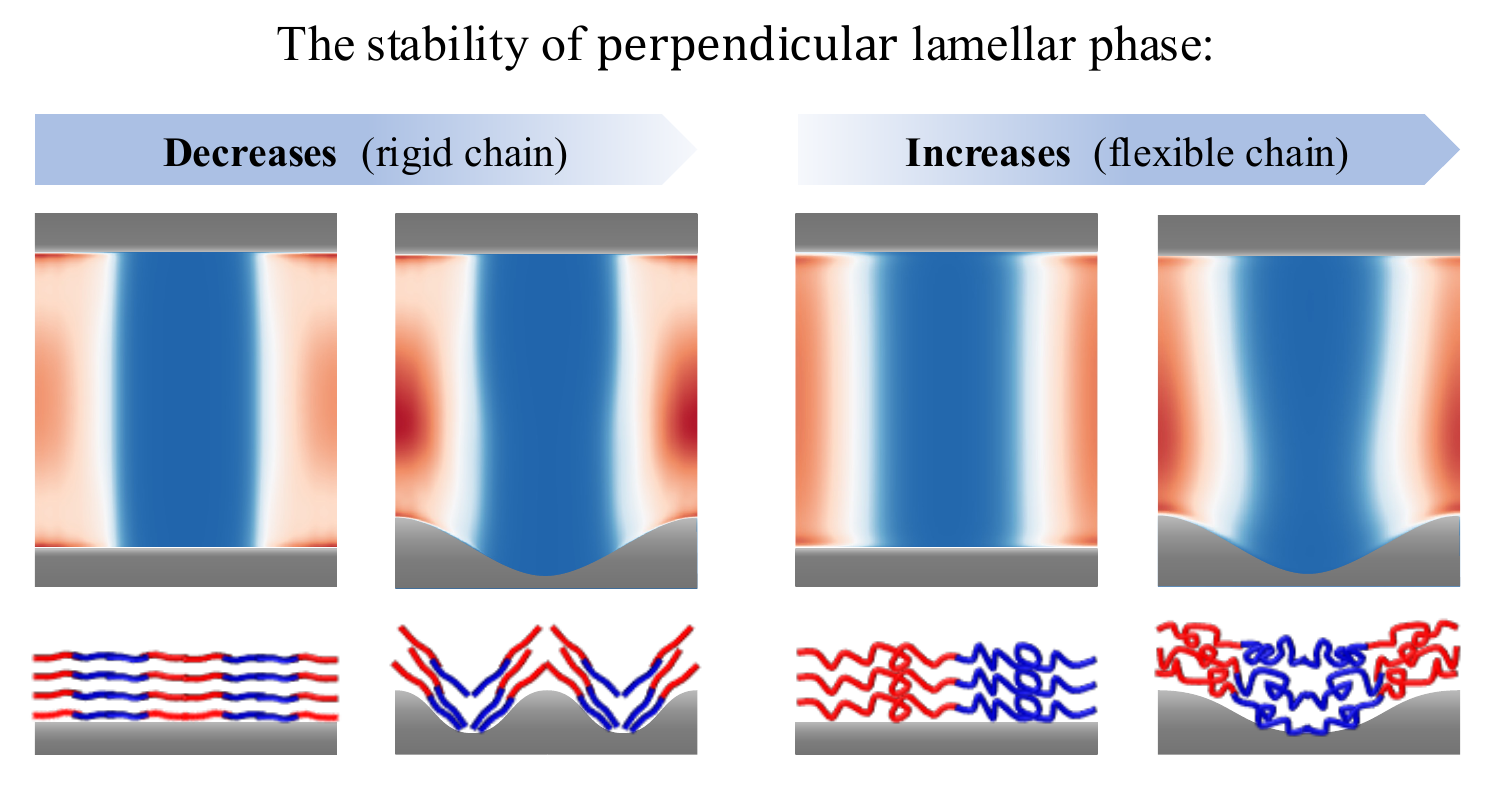}}
\end{figure}

\end{document}